\documentclass[11pt]{article}
\usepackage{emulateapj}
\usepackage{epsfig}

\def\Bruntfreq{Brunt-V{\"a}is{\"a}l{\"a}\,\,\,}
\def\refnew#1{(\ref{#1})}
\def\be{\begin{equation}}
\def\ee{\end{equation}}
\def\erg{\, \rm erg}
\def\K{\, \rm K}
\def\km{\, \rm km}
\def\s{\, \rm s}

\def\Hz{\, {\rm Hz}}

\def\kHz{\, {\rm kHz}}
\def\cm{\, {\rm cm}}
\def\yr{\, {\rm yr}}

\newcommand{\alphasat}{\alpha_{\rm sat}}
\newcommand{\alphacrit}{\alpha_{\rm crit}}
\newcommand{\Rey}{{\Re}}
\newcommand{\nukhz}{\nu_{{\rm kHz}}}

\begin{document} 

\title{R-modes in Neutron Stars with Crusts: Turbulent Saturation, 
Spin-down, and Crust Melting}
\author{Yanqin Wu, Christopher D. Matzner, and Phil Arras}
\affil{Canadian Institute for Theoretical Astrophysics, University of
Toronto}
\authoraddr{60 St George St, Toronto, Ontario M5S 3H8 Canada; wu, matzner, 
arras@cita.utoronto.ca}

\begin{abstract}
Rossby waves (r-modes) have been suggested as a means to regulate the
spin periods of young or accreting neutron stars, and also to produce
observable gravitational wave radiation. R-modes involve primarily
transverse, incompressive motions of the star's fluid core. However,
neutron stars gain crusts early in their lives: therefore, r-modes
also imply shear in the fluid beneath the crust. We examine the
criterion for this shear layer to become turbulent, and derive the
rate of dissipation in the turbulent regime. Unlike dissipation from a
viscous boundary layer, turbulent energy loss is nonlinear in mode
energy and can therefore cause the mode to saturate at amplitudes
typically much less than unity. This energy loss also reappears as
heat below the crust. We study the possibility of crust melting as
well as its potential implications for the spin evolution of low-mass
X-ray binaries.  Lastly, we identify some universal features of the
spin evolution that may have observational consequences.
\end{abstract}

\section{Introduction} \label{S:Intro}

The possibility of gravitational radiation from rapidly-rotating
neutron stars has recently come under intense scrutiny. This interest
has been fueled partly by the need to explain both the limited range
of rotation periods in low-mass X-ray binary systems (LMXBs) with
various accretion rates
\markcite{1998ApJ...501L..89B,1998ApJ...502..708A}({Bildsten} 1998; {Andersson} 1998) 
and the observed upper limit in spin rate of young neutron stars, and
partly by the possibility that the gravitational radiation from LMXBs
or newborn neutron stars may be detectable by gravity observatories
such as LIGO
\markcite{Owenetal98,1998ApJ...501L..89B,1999ApJ...516..307A}({Owen} {et~al.} 1998; {Bildsten} 1998; {Andersson}, {Kokkotas}, \&  {Stergioulas} 1999). 
For neutron stars with crusts, there are two potential sources of
gravitational radiation: a mass quadrupole originating in the solid
crust of the star \markcite{1998ApJ...501L..89B}({Bildsten} 1998), or a current
quadrupole from Rossby waves, which become overstable under the
influence of their own gravitational radiation
\markcite{1998ApJ...502..708A,1998ApJ...502..714F}({Andersson} 1998; {Friedman} \& {Morsink} 1998).

The outer layer of a neutron star solidifies if the typical energy of
Coloumb interactions between nuclei, $Z^2e^2/r$, exceeds the thermal
energy, $k_b T$, by a critical factor $172$
\markcite{1993PhRvE..47.4330F}({Farouki} \& {Hamaguchi} 1993). Here, $Z$ and $r$ are the mean charge and
the mean spacing of nuclei, respectively.  The crystallization (also
the melting) temperature is therefore
\begin{equation}
\label{eq:T9melt} 
T_{\rm melt} = 6.0\times 10^9 Z_{20}^2 A_{100}^{-1/3} \K,
\end{equation} 
where we have scaled $Z$ and $A$ (mean atomic weight) with values
suitable for the bottom of the LMXB crust \markcite{1990A&A...229..117H}
({Haensel} \& {Zdunik} 1990). Young neutron stars
\markcite{NegeleVautherin}({Negele} \& {Vautherin} 1973), and LMXBs whose crust
have been sufficiently heated,
may have higher values of $Z$ and therefore higher $T_{\rm melt}$
Nucleation of a crystalline crust begins at
around $T_{\rm melt}$, and crystallization accelerates exponentially
as the temperature is decreased further \markcite{1995ApJ...452..359D}({de Blasio} 1995).
All but the hottest neutron stars are expected to possess a solid
crust of thickness $\sim 1 \km$.
Our investigation continues along the
lines of \markcite{2000ApJ...529L..33B}{Bildsten} \& {Ushomirsky} (2000; hereafter, BU) to
explore the implications of such a solid crust for r-modes. We also
consider the fate of the crust during r-mode instability.

We shall focus on the fastest growing r-mode with spherical indexes
$\ell = m = 2$, whose rotating frame frequency is
$\omega=2\Omega_s/3$.  Here $\Omega_s$ is the spin frequency of the
neutron star; we denote its angular counterpart as $\nu_s \equiv
\Omega_s/2\pi$.  In the fluid core, the r-mode is largely horizontal, and the
 modulus of its displacement vector is \markcite{Owenetal98}({Owen} {et~al.} 1998)
\begin{equation}\label{eq:xi}
|\xi(r,\theta)| = \sqrt{\frac{45}{64\pi}} \alpha R \left(\frac{r}{R}\right)^2 
 \left[1-\cos^4\theta \right]^{1/2},
\end{equation}
where $\theta$ is the co-latitude,
$r$ the distance from the
center and $R$ the stellar radius. The amplitude of the velocity
perturbation is related to the displacement by $|v| = \omega
|\xi|$. If the neutron star crust is perfectly rigid, the r-mode
produces a periodic rubbing at the fluid-solid boundary with velocity
$|\Delta v| \approx |v(R)|$. However, \markcite{LU00}{Levin} \& {Ushomirsky} (2000) have argued that the
crust of a neutron star is not perfectly rigid, and it will
participate in the lateral motion of the mode.  To account for this,
we write $|\Delta v| = \eta |v|$, with $\eta = 1$ in the limit where
the shear force in the crust much exceeds the local Coriolis force. 

BU considered energy dissipation in a thin viscous boundary layer at
the above core-crust interface.  Dissipation due to ``molecular''
viscosity is more intense in this laminar boundary layer than in the
bulk flow caused by the mode. Therefore, the presence of the boundary
layer increases the stellar rotation rate above which the mode becomes
unstable, relative to the case when the crust is absent
\markcite{Owenetal98}({Owen} {et~al.} 1998).  
The mode is pumped by gravitational radiation and damped by
viscosity at rates that both scale as $\alpha^2$. So a laminar
boundary layer cannot provide a saturation mechanism for an unstable
r-mode.

In contrast, a turbulent boundary layer provides nonlinear dissipation
that leads to r-mode saturation.  A turbulent boundary layer occurs
once the mode amplitude grows above a critical value.  It removes
kinetic energy from the mode with a rate that is cubic in $\alpha$. As
$\alpha$ grows, turbulent dissipation rapidly catches up with the
amplification of the mode energy due to gravitational
back-reaction. We find this mechanism typically saturates the mode
amplitude $\alpha$ at values much less than unity, i.e., well below
what has been assumed in the literature
\markcite{1999ApJ...517..328L,Owenetal98}({Levin} 1999; {Owen} {et~al.} 1998)

The kinetic energy of the r-mode is converted into heat in the thin
turbulent boundary layer.\footnote{The heat input due to the viscous
boundary layer is insignificant except when 
the r-mode is marginally
unstable.}  If conduction and local neutrino emission cannot carry the
heat away sufficiently quickly, the local temperature may increase above
the melting temperature (eq. [\ref{eq:T9melt}]) and the crust will
begin to melt.  In \S \ref{sec:conduction}, we calculate the thermal
profile induced by boundary-layer heating at the core-crust
interface. 

Finally, in \S \ref{S:crust}, we consider the implications of
turbulent saturation and local heating on the thermal and spin
evolution of LMXBs as well as new-born neutron stars.
We conclude in \S \ref{S:conclusion}.

Throughout this paper we employ as a fiducial model an $n=1$ polytrope
of mass $M = 1.4 M_\odot$ and radius $R = 12.53 \km$
\markcite{LindblomMendellOwen}(as in {Lindblom}, {Mendell}, \&  {Owen} 1999). 
The bottom of the crust is assumed to have density $\rho = 1.5\times
10^{14}\, g\cm^{-3}$
\markcite{2000ApJ...531..988B}({Brown} 2000).

\section{Viscous and turbulent boundary layers}\label{S:explanation}

Going outward from the fluid core to the solid crust, the shear
modulus of the material changes from zero to a large (but finite)
value. This implies a discontinuity $\Delta v$ in the horizontal
velocity of the r-mode across the core-crust boundary.  The r-mode
periodically rubs the fluid core against the solid crust with this
velocity.  A viscous boundary layer develops for small velocity jump,
and a turbulent boundary layer sets in when the jump increases above
some critical value.

The effect of a viscous boundary layer on r-modes was first analyzed by
BU. We reiterate some of their results here.  Ignoring the Coriolis
force, the thickness of this layer is given by the diffusion length
during one mode period 
\markcite{Stokes51}({Stokes} 1851),
\begin{equation}\label{eq:diffusion}
\delta = \sqrt{{2\nu}/{\omega}},
\end{equation}
where $\nu$ is the molecular viscosity at the base of the crust, $\nu
\approx 1.8 \times 10^4 T_8^{-2}\cm^2\s^{-1}$ for the chosen density
\markcite{1979ApJ...230..847F,1987ApJ...314..234C}({Flowers} \& {Itoh} 1979; {Cutler} \& {Lindblom} 1987), 
and $T_8$ stands for $ T/10^8\K$. The rate of energy dissipation per
unit area is equal to the mean relative kinetic energy contained
within this layer (half of $\rho |\Delta v|^2/2$ times $\delta$)
multiplied by the mode frequency:
\begin{equation}
{{d\dot{E}_{\rm vbl}}\over{dA}} = - \frac{1}{4} \rho |\Delta v|^2 \cdot 
\delta \cdot \omega .
\label{eq:edot-vbl}
\end{equation}
This yields a global energy dissipation rate of 
\begin{equation} 
\dot{E}_{\rm vbl} \approx - 0.03 \s^{-1} E \eta^2 T_8^{-1}
\nukhz^{1/2},
\end{equation} 
where $\nukhz \equiv \nu_s/1\kHz$, and $E$ refers to the energy in the
mode \markcite{Owenetal98}({Owen} {et~al.} 1998),
\begin{equation}\label{eq:Emode}
E = \frac{1}{2} \alpha^2 \Omega_s^2 M R^2 {\tilde J},
\end{equation}
with ${\tilde J}= 0.016$.\footnote{This expression for $E$ assumes
that the r-mode satisfies equation \refnew{eq:xi} in the whole star;
strictly speaking, an impenetrable crust reduces the mode energy by a
factor of $\sim 2$.  We ignore this correction throughout the paper.}
BU found that the viscous boundary layer stabilizes r-modes against
gravitational back-reaction excitation at low spin frequencies and low
stellar temperatures.

When an unstable r-mode acquires sufficiently large amplitude, the
laminar boundary layer will make a transition to turbulence. This
requires the local Reynolds number to exceed a critical value
\begin{equation}\label{eq:critical}
\Rey \equiv {{|\Delta v| |\Delta \xi|}\over \nu} 
 \geq \Rey_{\rm crit},
\end{equation}
where the amplitude of the relative displacement 
$|\Delta \xi| = 
|\Delta v|/\omega$ and the critical Reynolds number is determined
experimentally to be $\Rey_{\rm crit} \sim 2\times 10^5$
\markcite{1989JFM...206..265J}({Jensen}, {Sumer}, \&  {Fredsoe} 1989).\footnote{Note that if we define a
Reynolds number using the width of the turbulent layer (as is done in
a pipe flow),
as opposed to using $\Delta \xi$,
 we find $\Rey_{\rm crit} \sim 500$ at the transition to
turbulence \markcite{1959flme.book.....L}({Landau} \& {Lifshitz} 1959).}

Combining the Reynolds number criterion (\ref{eq:critical}) with the
definition of $|\Delta \xi|$, we find that the onset of turbulence at
the equator requires a mode amplitude $\alpha$ greater than a critical
value $\alphacrit$:
\begin{equation}
\alpha_{\rm crit} = 1.6\times 10^{-3} \eta^{-1} T_8^{-1} \nukhz^{-1/2},
\label{eq:alpha_crit}
\end{equation}
which corresponds to a thickness of the boundary layer $d = 31 \cm
T_8^{-1} \nukhz^{-1}$ (see Appendix \ref{S:shearturb}).  At other
latitudes, turbulence sets in at a larger mode amplitude.

Turbulence implies that eddies, rather than viscosity, advect momentum
toward the crust.  The crust feels a drag force very nearly in phase
with $\Delta v$ with a magnitude (eq. [\ref{eq:dragold}])
\begin{equation}
{\rm Drag} \equiv C_D\cdot \rho |\Delta v|^2/2,
\label{eq:drag}
\end{equation}
where the drag coefficient $C_D$ is measured to be $C_D \sim 5\times
10^{-3}$ \markcite{1989JFM...206..265J}({Jensen} {et~al.} 1989) at 
the onset of turbulence and varies logarithmically with the Reynolds
number beyond the critical point, as is discussed in Appendix
\ref{S:shearturb}. Kinetic energy is removed from the r-mode with a
time-averaged rate per unit area of
\begin{equation}
{{d\dot{E}_{\rm turb}}\over{dA}} = - {1\over 2} |\Delta v|\cdot |{\rm Drag}|,
\label{eq:eturbdA}
\end{equation}
where time-averaging gives the factor $1/2$ as the drag is in phase
with $\Delta v$.
Integrating over the sphere assuming a constant $C_D$, we find the
total dissipation rate
\begin{equation}
\dot{E}_{\rm turb} \simeq -{{3\pi}\over 4} \rho R^2 C_D |\Delta v_1|^3,
\label{eq:turbdEdt}
\end{equation}
where $|\Delta v_1| = (5/16\pi)^{1/2} \alpha \Omega_s R$ is the
velocity jump at the equator. In general, we adopt for $C_D$ the value
at the equator, where the Reynolds number is the largest.

Turbulent dissipation increases with $\alpha$ more rapidly than does
the pumping of the mode by gravitational back-reaction, and this leads
to saturation. Recall that the mode gains energy due to gravitational
wave radiation at the rate
\markcite{Owenetal98}(cf. {Owen} {et~al.} 1998)
\begin{equation}
\dot{E}_{\rm gr} = 0.11 \s^{-1} \, E\, \nukhz^6.
\label{eq:grdEdt}
\end{equation}
Balancing this energy gain with the nonlinear energy drain from
turbulence, we find a saturation amplitude
\begin{equation}
\alphasat \simeq 3.5 \times 10^{-3} \, \eta^{-3} \nukhz^5 
\left({{5\times 10^{-3}}\over C_D}\right).
\label{eq:alpha_satu}
\end{equation}
Note that $\alphasat$ is independent of temperature up to the
logarithmic dependence of $C_D$ on $\Rey$. Typically $\alphasat \geq
\alpha_{\rm crit}$ for an unstable r-mode, in which case the r-mode
grows rapidly until it reaches a steady state with $\alpha =
\alphasat$.  The scaling $\alphasat \propto \eta^{-3} \nukhz^{5}$
seems to imply that the saturation amplitude will increase enormously
(at a given spin frequency) if the crust's rigidity parameter $\eta$
is reduced from unity to $0.1$.  Newborn, rapidly spinning neutron
stars will therefore spin down at a rate that is very sensitive to
$\eta$ (and may be saturated by other means if 
eq. [\ref{eq:alpha_satu}] predicts $\alphasat > 1$). In contrast, 
the spins of LMXBs are confined within a narrow frequency range that
varies with $\eta$ (as we will show in \S \ref{S:crust}). 
The spin frequency at which
they first become unstable scales as $\nu_s
\propto \eta^{4/11}$ (equating eqs.[\ref{eq:edot-vbl}] with
[\ref{eq:grdEdt}]). 
Taking this change of spin frequency into account,
we find that the maximum r-mode amplitude in LMXBs varies as
$\eta^{-1.2}$. Low saturation amplitudes are expected in LMXBs even
for $\eta\sim 0.1$.

We 
return now to comment on a few 
issues relevant to turbulence
onset. The first 
is related to the local stratification. We find
the Richardson number
\begin{equation}\label{eq:Richardson}
{\cal J} \equiv {{N^2}\over{\left|{{d v}/{dr}}\right|^2}} \sim C_D
N^2/\Omega_s^2 \ll {1\over 4}
\end{equation}
at $\alpha = \alphasat$ for a \Bruntfreq frequency of $N\sim 500\ {\rm rad}\
\s^{-1}$ \markcite{1992ApJ...395..240R}({Reisenegger} \& {Goldreich} 1992). This indicates
that
the stable stratification at the core-crust boundary does not prevent
the onset of turbulence.

Notice that the Reynolds number (eq. [\ref{eq:critical}]) can also be
written as $\Rey = 4 (|\Delta \xi|/\delta)^2 \simeq 2 C_D
(d/\delta)^2$, where the thickness of the turbulent layer $d$ is
related to $|\Delta \xi|$ as $d \simeq (C_D/2)^{1/2} |\Delta \xi|$
(see Appendix \ref{S:shearturb}). The condition $\Rey \geq
\Rey_{\rm crit}$ implies $d \gg \delta$. Viscosity is unimportant in 
most of the turbulent region. Moreover, the turn-over time of the
energy-bearing (also the largest) eddies in the turbulent region is
much shorter than the mode period, 
$|\Delta v|/d \gg \omega$. This allows for
a well-developed turbulent cascade.

The interaction between shear turbulence and stellar rotation, the
possibility of equipartition magnetic fields in the turbulent layer,
and the possible superfluid nature of the material, are potentially
important questions that lie beyond the scope of this paper.

\section{Temperature Profile from Boundary Layer Heating}
\label{sec:conduction}

Turbulence converts the kinetic energy of the r-mode into heat and
deposits it in a thin region of width $d \simeq 100
\cm \eta^{-2} \nukhz^5 $ below the crust.  If thermal conduction and
local neutrino emission cannot carry away the heat efficiently,
the
temperature in this layer may be driven up to past the melting
temperature of the crust (eq. [\ref{eq:T9melt}]).  As our saturation
mechanism (\S \ref{S:explanation}) depends on the presence of such a
solid crust, we wish to establish the temperature profile of the
boundary layer. 
We assume a steady state solution because the conduction time over the
regions of interest to us is much shorter than the evolution time. 

The conduction of heat away from the boundary layer requires 
a temperature profile that declines in each direction.
This, in turn, implies that
the rate of local neutrino cooling decreases with distance away from
the boundary layer. We assume that the scale length of the temperature
perturbation is much shorter than the density scale height, and
therefore much smaller than the stellar radius as well. This
assumption, whose validity must be checked {\em a posteriori}, allows
us to consider only planar heat conduction in a region of constant
density.

The following toy model gives insight into the full solution. Let us
consider a piecewise linear temperature profile, one that reaches a
temperature $T_{\rm bl}$ within the boundary layer (of width $d$), and falls to zero
over lengths
$l_{\rm core}$ and $l_{\rm crust}$ to the inside and outside,
respectively. The two 
lengths differ because the thermal
conductivity takes different values in the two directions.

In a state of equilibrium, any energy not radiated as neutrinos within
the boundary layer must be carried away by conductive fluxes:
\begin{equation}\label{eq:conservation}
{{d\dot{E}_{\rm turb}}\over{dA}} = d\cdot \epsilon_0 \, T_{\rm bl}^8 + F_{\rm
core} + F_{\rm crust},
\end{equation}
where 
neutrino emissivity due to modified Urca reactions is $\epsilon_\nu
\approx \epsilon_0 T^8 \erg
\cm^{-3}\s^{-1}$ with $\epsilon_0 = 7.4\times 10^{-52}$
\markcite{1979ApJ...232..541F}({Friman} \& {Maxwell} 1979).  The heat flux leaving the boundary layer
in each direction is
\begin{equation}\label{eq:conductiveflux}
F_{\rm core} =
\kappa_{\rm core}\, \frac{T_{\rm bl} }{l_{\rm core}}
 {~~\rm and~~ } F_{\rm crust} = 
\kappa_{\rm crust}\, \frac{T_{\rm bl}} {l_{\rm crust}},
\end{equation} 
respectively. The thermal conductivity of the crystalline crust is
dominated by electron-phonon scattering and scales with the local
temperature inversely, $\kappa_{\rm crust} =
\bar{\kappa}_{\rm crust} T_8^{-1} \approx 
10^{20}T_8^{-1} \erg \cm^{-1}\s^{-1}\K^{-1}$,
while that of the liquid core is dominated by electron-proton and
electron-neutron scattering with the same temperature dependence,
$\kappa_{\rm core} =
\bar{\kappa}_{\rm core} T_8^{-1} \approx 10^{23}T_8^{-1} \erg
\cm^{-1}\s^{-1}\K^{-1}$ \markcite{1979ApJ...230..847F}({Flowers} \& {Itoh} 1979). 

Another consequence of thermal equilibrium is that 
the heat conducted into the crust or core must be radiated there in
neutrinos. With our approximation of a linear temperature profile,
this implies $F \approx
\epsilon_0 l T_{\rm bl}^8/9$ in each direction with the conductive lengths
$l$ defined in equation \refnew{eq:conductiveflux}. A more
accurate calculation 
in which no temperature profile is assumed {\it a priori}
(Appendix \ref{S:ConductionCalc})
gives $F = \epsilon_0 l T_{\rm bl}^8/4$. Combining this with equation
\refnew{eq:conductiveflux}, we find the conductive lengths 
to be
\begin{eqnarray}\label{eq:lengths}
l_{\rm core} & \simeq & 2.4\times 10^5 \cm \, 
\left({{10^9\K}/{T_{\rm bl}}}\right)^{4}, \nonumber \\
l_{\rm crust} & \simeq & 7.4\times 10^3\cm  \,
\left({{10^9 \K}/{T_{\rm bl}}}\right)^{4}.
\label{eq:conduction_length}
\end{eqnarray}
As long as $T_{\rm bl} \geq 5\times 10^8 \K$, $l_{\rm crust} < l_{\rm
core} \ll R$ and our assumption of a thin heated region is
justified. The heating becomes increasingly local as the boundary
layer rises in temperature: when more heat is deposited in the
boundary layer, a larger temperature gradient is required to conduct
the heat outward leading to a smaller conduction length. The
assumption of an isothermal star is thus not justified in the presence
of any significant localized heating. In \S \ref{subsec:isothermal}
and \S \ref{subsec:realistic} we shall compare the implications of the
isothermal approximation with those of localized temperature profile
for neutron star spin evolution.

Substituting equations \refnew{eq:conductiveflux} and
\refnew{eq:conduction_length} into equation \refnew{eq:conservation}, 
adopting $\alpha=\alphasat$ and 
evaluating $d\dot{E}_{\rm turb}/dA$ at the equator,
we find a quadratic equation for $T_{\rm bl}^4$,
whose solution gives
\begin{equation}
T_{\rm bl} =  1.5\times 10^{10} \K\, \eta^{-1/2}\, \nukhz^{13/8} 
\left( \sqrt{1+b^2}-b \right)^{1/4},
\label{eq:T_bl}
\end{equation}
where $b=7.4 \times 10^{-3} \eta^4 \nukhz^{-23/2}$ and $C_D$ is taken
to be $0.005$. Neutrino losses from the boundary layer itself
dominate over the conductive fluxes
for $b \leq 1$, or $\nu_s \geq 650
\Hz\, \eta^{8/23}$. For a given mode amplitude, the boundary layer 
temperature
decreases towards the poles.

We find that the flux going to the core is roughly $(\kappa_{\rm
core}/\kappa_{\rm crust})^{1/2}\sim 30$ larger than that conducted to
the crust. However, if the above equilibrium temperature exceeds
$T_{\rm melt}$, the crust will begin to melt.  Once melting begins, we
expect that the energy input from the boundary layer will be used
almost entirely to melt the crust. The temperature in this layer will
be kept very close to $T_{\rm melt}$ by the continual inflow of cool,
freshly molten crust, as the boundary layer moves upward following the
bottom of the receding crust. The time it takes to completely melt the
crust can be estimated to be a few days (the ratio
$10^{48}\erg/\dot{E}_{\rm gr}$ where $10^{48} \erg$ is roughly the
chemical binding energy of the crust).

\section{Implications for Neutron Star Spin and Temperature Evolution}
\label{S:crust}

We now consider the implications of turbulent saturation for the
evolution of young neutron stars and LMXBs in the $(\Omega_s, T)$
plane.  In this section we assume the crust is present and is not
melted, an assumption we examine in \S
\ref{subsec:realistic}. Once the neutron star is pushed into the
instability region, either by cooling or by accretion, its r-mode
quickly grows to its saturated value as given by
equation \refnew{eq:alpha_satu}. 
Inserting
it into equation
\refnew{eq:omegadot} we find that the time to spin down to
a frequency $\nukhz$ is fairly independent of the initial spin
frequency of the star, and is given by
\begin{equation}
t  =  5.2 \times 10^5 \s \,\eta^6 \nukhz^{-16}.
\label{eq:spindowntime}
\end{equation}
The time to spin the star down to, say, $\nu_s=500 \Hz$ is roughly
$10^3 \eta^6 \yr $. 
As $\alphasat$ decreases sharply with decreasing frequency, most of
the spin-down time is spent near the lowest frequencies. 

The above conclusion on the evolution of the spin frequency may need
modification if the crust can be melted by the r-mode heating. This
prompts us to study the temperature evolution of the star.  This we do
by first adopting the isothermal approximation for the whole star. It
applies in the limit of infinite thermal conductivity or large-scale
heating.  We then apply our results from \S \ref{sec:conduction} to
the realistic case in which the conductivity is finite and r-mode
heating is local.

Evolution in the unrealistic isothermal case can be easily integrated
using equations in Appendix \ref{S:evolutioneqn}. It is useful for
comparison with previous work. More importantly, insights gained from
the isothermal case can be applied to the realistic case.

\subsection{Isothermal Approximation}
\label{subsec:isothermal}

\begin{figure*}
\centerline{\psfig{figure=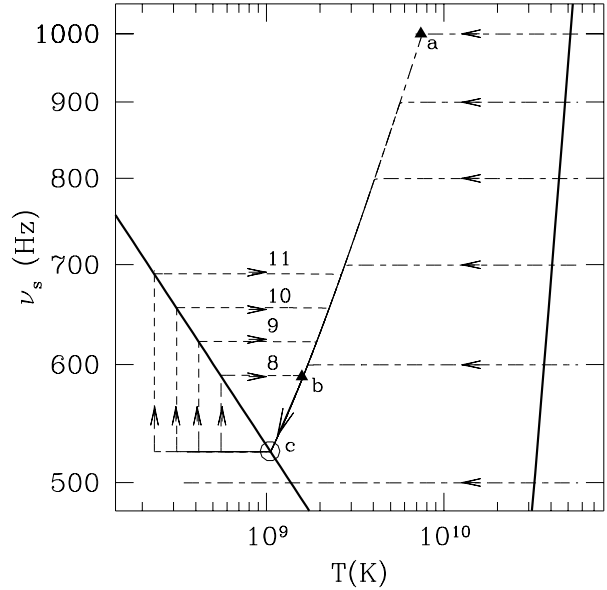,width=0.45\hsize}}
\caption[]{Evolutionary tracks for LMXBs and young neutron stars under
the isothermal approximation, with arrows indicating the
direction. The {\it thick lines} extending up from the bottom enclose
the
r-mode instability region.  The {\it dashed loops} labeled $8-11$ are
LMXB tracks for accretion rates $\dot{M}/M=10^{-8}-10^{-11}
\yr^{-1}$. The 
{\it long-dashed-short-dashed curves} are young neutron star
tracks with a range of initial frequencies. Note that all tracks that
enter the instability region above $\nu_T \simeq 520 \Hz$ spin down
along the equilibrium line (eq. [\ref{eq:coolspin}]) and exit at
$\nu_s = \nu_T$ (labeled as $c$). Below this frequency the cooling
time is much shorter than the spin-down time
(eq. [\ref{eq:spindowntime}]). Symbols $a$ and $b$ mark the place
where two specific tracks first hit the equilibrium line. See Figure
\ref{fig:time} for more details on these two tracks.}
\label{fig:both}
\end{figure*}

\begin{figure*}
\centerline{\psfig{figure=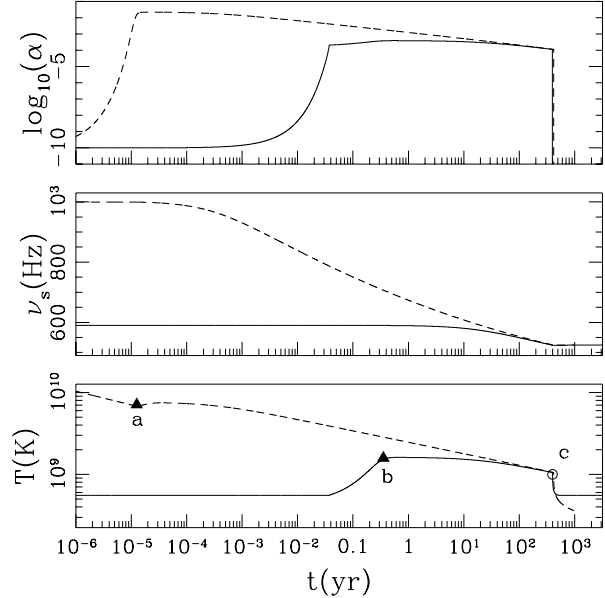,width=0.45\hsize}}
\caption[]{
R-mode amplitude $\alpha$, stellar spin frequency $\nu_s$ and temperature $T$
are plotted here as functions of time along
two evolutionary tracks: a young neutron star with an initial
frequency $\nu_s = 1 \kHz$ ({\it dashed lines}) and an LMXB with
$\dot{M}/M=10^{-8} \yr^{-1}$ 
({\it solid lines}).  The r-mode growth
rate in the case of the LMXB does not become appreciable until
accretion has pushed the star sufficiently far into the instability
region. Both tracks exit the instability region at the same spin
frequency ($\nu_T$) after spending roughly the same amount of time
(eq. [\ref{eq:spindowntime}]). Symbols $a$ and $b$ mark the points
when r-mode heating is first balanced by neutrino cooling and the
stars 
begin
to evolve along the equilibrium spin down line. This
occurs after r-modes have reached the saturation amplitudes.}
\label{fig:time}
\end{figure*}

After the r-mode reaches its saturation amplitude, it provides the
neutron star with a 
source of heat from turbulent
dissipation. 
The neutron star heats up quickly until it reaches a state in which
the r-mode heating is balanced by the neutrino losses. With time the
temperature will decrease slowly because heating by r-mode decreases
with the spin frequency.
We set the r-mode heating term equal to the neutrino cooling
term in equation \refnew{eq:Tdot} to find the following scaling
between the temperature and the spin rate of the star,
\begin{equation}
T_8 = 47 \eta^{-3/4} \nukhz^{9/4},
\label{eq:coolspin}
\end{equation}
where we take $\alpha = \alphasat$ and ignore the logarithmic
dependence of $C_D$ on temperature by using $C_D = 0.005$. This
relation gives us the thermal equilibrium spin-down line.  It
intersects the instability curve at the {\em terminal frequency}
$\nu_T \simeq 520 \Hz\ \eta^{0.35}$.  {\it All} neutron stars entering
the instability region with an initial spin frequency $\nu_s \geq
\nu_T$ will converge onto the equilibrium line and exit at
$\nu_T$. The ones that enter at $\nu_s < \nu_T$ find their spin rates
hardly affected by r-mode instability. This is because their spin down
time (eq. [\ref{eq:spindowntime}]) much exceeds their neutrino cooling
time.

Figures \ref{fig:both} and \ref{fig:time} exhibit the results of our
evolutionary calculations with a variety of mass accretion rates in
LMXBs and a range of initial spin rates for young neutron stars.  The
evolution equations we adopted for the integrations are listed in
Appendix \ref{S:evolutioneqn}. They are similar to those derived in
\markcite{Owenetal98}{Owen} {et~al.} (1998) apart from 
the additional terms that arise from boundary layer dissipation.
We consider only $\eta = 1$ here. We find that, up to logarithmic
dependence of $C_D$ on the Reynolds number, the numerical results
confirm equations \refnew{eq:spindowntime} and \refnew{eq:coolspin}.

Note that the presence of an equilibrium spin-down line is a universal
feature independent of the saturation mechanism. If r-modes are
saturated to constant values, for instance, we find the equilibrium
temperature $T \propto \nu_s$.

\markcite{1999ApJ...517..328L}{Levin} (1999) was the first to point out that LMXBs
undergo limit cycles in which the angular momentum accreted over the
whole cycle is radiated by gravitational waves when the r-mode is
unstable.  The r-mode phase lasts a time that depends on the
saturation mechanism and is relatively insensitive to the initial spin
frequency at which the r-mode is destabilized. Equation
\refnew{eq:spindowntime} yields the spin-down time at the terminal
frequency $\nu_T$ to be $\sim 10^3
\eta^{10/31} \yr$. Compared to an accretion spin-up time of 
$\sim 10^7 \yr$ at the Eddington accretion rate, we find a maximum
duty cycle of $\sim 10^{-4}$ during which time the gravitational wave
emitted by the r-mode may be potentially detectable, with the duty
cycle decreasing at lower accretion rate.  This duty cycle is much
higher than that found by previous investigators
\markcite{1999ApJ...517..328L}(e.g., {Levin} 1999) mostly due to our much reduced
saturation amplitude.

\subsection{Realistic Evolution Accounting for Local Heating} \label{subsec:realistic}

\begin{figure*}
\centerline{\psfig{figure=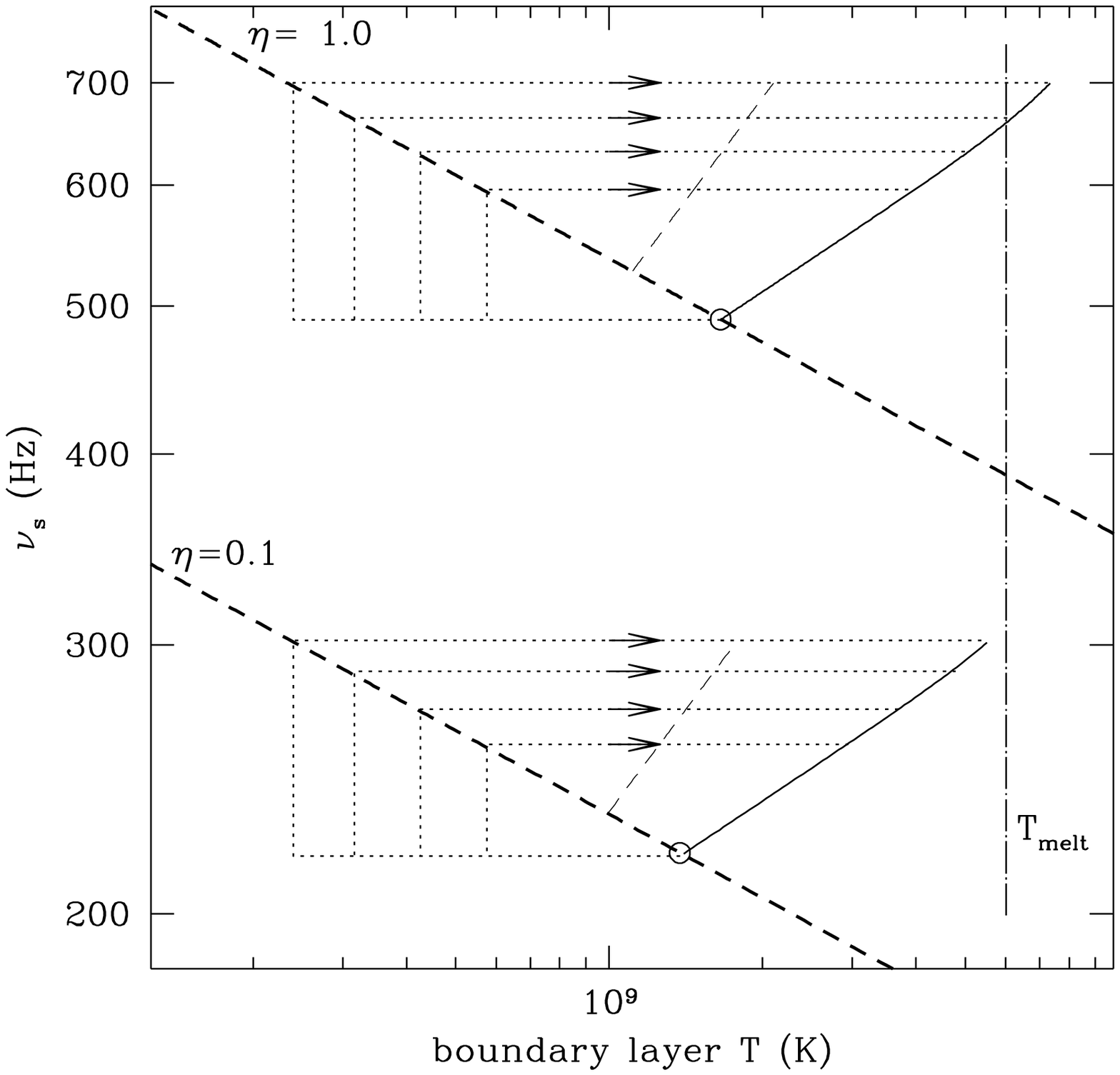,width=0.45\hsize}}
\caption[]{Limit cycles for the same four LMXBs in Figure \ref{fig:both}
are plotted as {\it dotted loops} against the temperature at the
boundary layer. The {\it heavy dashed curves} delineate
the instability regions and the {\it slanted solid lines} show
the equilibrium spin-down curves (eq. [\ref{eq:T_bl}]) for two
different values of $\eta$.  The large circles
denote $\nu_T$ for each case.  The {\it light dashed lines} are drawn
for comparison -- they are equilibrium spin-down curves under the
isothermal approximation.  We also 
show
the melting line ({\it
dot-dashed}) with $T_{\rm melt} = 6\times 10^9
\K$.}
\label{fig:conduction}
\end{figure*}

For simplicity, we consider only the evolution of LMXBs in this
section. Numerical integration for the evolution tracks in the
realistic case is 
much more difficult than in the isothermal
case. Fortunately, they are not necessary. As is evident from Figure
\ref{fig:both}, given initial conditions, the equilibrium line and the
instability curve completely determine the evolution tracks. 

The r-mode instability curve in the low temperature range depends only
on the boundary layer temperature because damping from the viscous
boundary layer dominates over other damping mechanisms. The
equilibrium spin-down line 
is also given in terms of the boundary layer temperature
(eq. [\ref{eq:T_bl}]). So we can simplify the problem in hand by
considering only the temperature at the boundary layer.

There are two main differences between the realistic and the
isothermal case: a lower terminal frequency and an enhanced heating in
the boundary layer.

The new equilibrium spin-down line (eq. [\ref{eq:T_bl}]) intersects
the instability curve at a terminal frequency
\begin{equation}
\nu_T \simeq 490 \eta^{0.35}\, \Hz
\label{eq:newterminal}
\end{equation}
that is {\it lower} than in the isothermal case. It takes slightly
longer to spin the star down to this lower value of $\nu_T$.

The r-mode heats the boundary layer to a higher temperature than under
the isothermal approximation. This results from the fact that, unlike
in the isothermal case, there is now only a shell of thickness
$(l_{\rm core}+l_{\rm crust})/4 + d$ which receives heat and cools by
neutrinos (\S \ref{sec:conduction}).
 
We present the LMXB limit cycles for the two cases $\eta = 1$ and
$\eta = 0.1$ in Figure \ref{fig:conduction}.  One observational
consequence suggests itself in the figure: all LMXBs are expected to
have spin frequencies falling in the narrow range of $490 - 700 \Hz$
when $\eta = 1$, and $220 - 300 \Hz$ when $\eta = 0.1$, independent of
their accretion history. This continues the idea proposed by
\markcite{1998ApJ...501L..89B}{Bildsten} (1998) and
\markcite{1999ApJ...516..307A}{Andersson} {et~al.} (1999) that r-modes
may be instrumental in halting the LMXB spin-up. The resemblance
between the above frequency range for $\eta = 0.1$ and the observed
LMXB spin rates \markcite{VanDerKlisARAA}({van der Klis} 2000)
is intriguing.
However, it  may be difficult to explain the fastest millisecond pulsars
(period of $1.5$ ms) using the same value of $\eta$.

The above results may need modification if the crust melts during the
r-mode evolution.  Melting the crust is more likely than 
is suggested by the isothermal approximation.
As Figure \ref{fig:conduction}
demonstrates, the crust is most likely to melt if $\eta$ is large and
if the accretion rate is low.  We define a melting frequency, $\nu_m$,
at which the equilibrium line intersects the melting line $T = T_{\rm
melt}$. We find for $T_{\rm melt} = 6\times 10^9 \K$, $\nu_m = 660
\Hz$ for $\eta = 1$ and $315
\Hz$ for $\eta = 0.1$.  We speculate 
below on some possible evolutionary consequences 
when crust melting is taken into account.

\subsection{Possible Outcomes Including Crust Melting and Forming}
\label{subsec:melt}

\begin{figure*}
\centerline{\psfig{figure=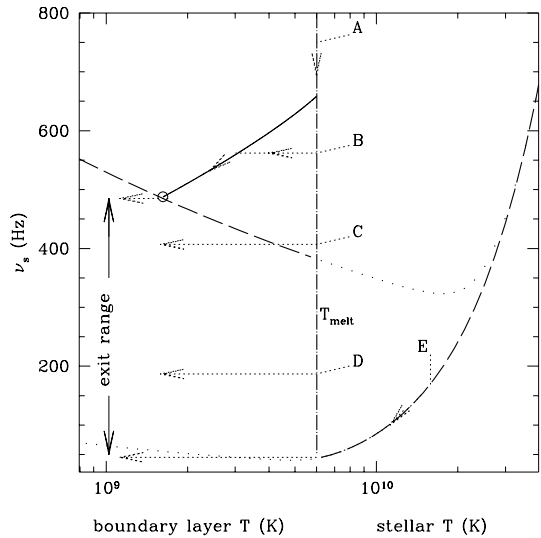,width=0.45\hsize}}
\caption[]{Possible outcomes of spin evolution when we consider crust 
melting and forming.  The horizontal axis represents the temperature
of the boundary layer as long as this is less than the melting
temperature (the {\em dot-dashed line}, we take $T_{\rm melt} =
6\times 10^9 \K$ as is appropriate for nuclei with $Z=20$); for higher
temperatures this axis represents the temperature of the fluid
core. The two {\em dashed lines} are instability curves that are valid
in the presence (to the left
of the melting line) or absence (to the right) of a crust,
respectively. Equation \refnew{eq:T_bl} gives the equilibrium,
saturated spin-down behavior ({\em solid line}) when local heating is
considered. We demonstrate the evolution ({\em dotted lines with
arrows}) for five scenarios in which the crust forms at different
initial spin frequencies. We find their final rotation rates reside in
the range marked as the `exit range'. In this figure, we have taken
$\eta = 1$ for illustrative purposes.}
\label{fig:evolution}
\end{figure*}

The rotational energy ($\sim 10^{51}\nukhz^2\erg$) far exceeds the
chemical binding energy of the crust ($\sim 10^{48}\erg$) for spin
rates of interest: crust melting is not inhibited by lack of energy.
Moreover, as discussed in \S \ref{sec:conduction}, the temperature at
the bottom of the crust may be driven above the melting temperature by
turbulent heating; this causes melting. For our fiducial parameters,
this occurs for slowly accreting LMXBs if $\eta = 1$
(Fig. \ref{fig:conduction}).

After the crust is melted, some other nonlinear dissipation must cause
the r-mode to saturate. It is beyond the scope of this paper to
predict the evolution when the crust is molten.  However, we can be
confident that the star will cool and spin down sufficiently to form a
crust again.  Once this occurs, the subsequent evolution is again
within our jurisdiction, and it depends on the stellar spin frequency
($\nu_e$) at the moment the crust forms.  We discuss five
representative evolution scenarios that begin with different values of
$\nu_e$, labeled A through E in Figure \ref{fig:evolution}.

In case A, the stellar temperature drops below the melting temperature
while the star is still spinning so fast ($\nu_e > \nu_m$) that the
boundary layer temperature would exceed $T_{\rm melt}$ if a crust {\em
did} form.  Therefore, the material that could form a crust is kept
hot (by some means) with $T \sim T_{\rm melt}$.  This implies that the
formation of an appreciable crust is delayed until the rotation has
slowed to $\nu_s = \nu_m$, where $\nu_m$ is the frequency at which the
equilibrium line hits the melting line.  After this point the star
will spin down with a saturated r-mode (eq. [\ref{eq:spindowntime}])
and cool in a state of thermal equilibrium (eq. [\ref{eq:T_bl}]); this
behavior is represented by the solid line in Figure
\ref{fig:evolution}. The star leaves the instability region with a
specific (terminal) spin frequency when the mode is no longer
unstable.

Case B represents the outcome if the initial spin frequency $\nu_e$ is
low enough to allow crust formation, yet higher than the terminal spin
frequency.  A crust forms and the star evolves towards the equilibrium
spin-down line, with the the cooling of the boundary layer regulated
by that of the whole star.  The subsequent evolution is identical to
case A.

Case C enters with $\nu_s < \nu_T$. Its spin-down time is longer than
the cooling time and it exits the r-mode instability curve without
being significantly spun down.  Case D has a similar outcome: r-mode
instability is quenched entirely once the crust forms.

Case E hits the instability curve to the right of the melting line. It
evolves along the critical stability curve until the melting line,
after which the r-mode becomes stable. This is the only scenario in
which we can predict the evolution of an r-mode in the absence of a
crust.

From these considerations we conclude that the final rotation rates of
young neutron stars, and of those LMXBs capable of melting their
crusts, depend on their spin evolution while the crust is absent.
In the case that $\eta = 1$ and $T_{\rm melt} = 6\times 10^9
\K$, the final spin frequencies should lie in the range of $50$ to $480
\Hz$,  where the upper limit is the terminal frequency $\nu_T$.
The chosen value of $\eta$ should be considered uncertain, but recent
calculations \markcite{LU00}({Levin} \& {Ushomirsky} 2000) point to
$\eta \sim 0.1$. Also, a minimum value of $\eta \sim 0.003$ can be
derived from equation \refnew{eq:newterminal} by assuming that $\nu_T$
is larger than the frequency of the fastest known young neutron star
($\sim 16 {\rm ms}$; \markcite{1998ApJ...499L.179M}{Marshall} {et~al.}
1998).
Moreover, young neutron stars, or LMXBs that have been
heated above the temperature necessary for nuclear burning, are more
likely to be in a state of chemical equilibrium at the base of their
crusts; this implies higher values of $Z$, thus higher melting
temperatures. However, the above discussion demonstrates that this
would only raise the lower bound for the possible range of final
rotation rates without changing the upper bound ($\nu_T$).

\section{Conclusions}
\label{S:conclusion}

R-modes in the fluid core of a neutron star generate velocity shear
below the solid crust.  The primary conclusion of this paper is that
this velocity shear can drive turbulence which in turns limits the
amplitude that an overstable r-mode can obtain. This saturation
amplitude is easily calculable. It rises steeply with mode frequency
and depends on the temperature at the turbulent boundary layer only
logarithmically. For most spin frequencies of interest, we find this
amplitude falls well below unity.

We also find that heating due to turbulent dissipation in the boundary
layer leads to a localized temperature enhancement relative to the
surrounding material, and the heat is removed by neutrinos from both
the boundary layer itself and from the nearby regions to which the
heat is conducted.  It is erroneous to assume that the whole star
remains isothermal in the presence of a localized heat source.  The
boundary layer temperature may exceed the melting temperature of the
crust.

When studying the spin and temperature evolution of neutron stars
undergoing r-mode instability, we find the following remarkable
behaviors: because of the strong dependence of neutrino cooling on the
temperature, the temperature of the boundary layer rapidly reaches an
equilibrium value that depends on the mode amplitude and spin
frequency, and hence, for a saturated mode, only on the stellar
spin. This defines an equilibrium spin-down line along which all
neutron stars evolve. We define the spin frequency at which this line
intersects the instability line as $\nu_T$, the terminal frequency.
This frequency is a significant feature of the spin evolution. We
expect all neutron stars with crusts to exit the r-mode instability
range with $\nu_s = \nu_T$, as long as their initial spin frequencies
are higher than $\nu_T$.  Neutron stars born with $\nu_s < \nu_T$ will
never experience significant spin-down. If the crust can be melted and
reformed, the terminal frequency defines the upper limit to the final
spin rates of these neutron stars.  We also establish a spin-down law
for which the time required to spin the star down to $\nu_T$ is
roughly independent of the initial spin rate.  These features are
expected to persist for other saturation mechanisms.

This work has relied on several assumptions and approximations which
should be noted. We have ignored any dissipation that may occur within
the solid crust itself, the possibility that the crust may be melted
by mechanical strain or heat from nuclear reactions, and additionally
any nonlinear dissipation caused by coupling between modes in the bulk
of the fluid core \markcite{Cornell}({Schenk} {et~al.} 2000). We have
ignored the possibility that gravitational radiation may be dominated
by a permanent mass quadrupole within the crust.  We have not
considered the effects of superfluidity in the material below the
crust.
 The detailed value of the boundary layer temperature depends
on the exact value of the thermal conductivity as well as on the
parameter $\eta$ that describes the crust motion; these we consider
relatively uncertain. Moreover, melting of the crust is sensitive to
the composition at its base, and this may differ between young and
accreting neutron stars.  Finally, our discussions in \S
\ref{subsec:melt} do not include the possibility that chemical
composition in the crust may change when melting occurs.

\acknowledgements We wish to thank Peter Goldreich, Norm Murray, Greg
Ushomirsky and Marten van Kerkwijk for comments and encouragements.
Each of the authors received support from an NSERC fellowship. CDM was
also supported by a visiting postdoctoral fellowship at Caltech; he
wishes to thank Sterl Phinney and Roger Blandford for their gracious
hospitality during this visit.  PA wishes to thank Katrin Schenk, Ira
Wasserman, Eanna Flanagan, and Saul Teukolsky for useful
conversations.


\appendix

\section{Turbulent Boundary Layer}
\label{S:shearturb}

In this appendix, we give a short description of the turbulent
boundary layer caused by shear flows. We provide scalings for the
thickness of the layer, the velocity profile inside the layer and the
drag force acting on the wall. These scalings are confirmed by
numerical experiments and we use them in the main text to derive the
energy dissipation rate due to shear turbulence.

Let $\Delta v$ be the oscillatory rubbing velocity between the neutron
star's fluid core and its solid crust. The horizontal displacement is
$\Delta \xi=\Delta v/\omega$.  When the boundary layer becomes
turbulent, we can describe it with results from a similar situation:
the turbulent boundary layer generated by a steady flow along a
semi-infinite plane.
Our discussion therefore follows closely that of
\markcite{1959flme.book.....L}{Landau} \& {Lifshitz} (1959, \S 42-44). 
In this case, $\Delta \xi$ can be understood as the distance away from
the edge of the plane, and $\Delta v$ as the velocity of the shear
flow.  Although the steady solution is expected to reproduce the
correct scalings for the boundary layer structure, it is
time-independent and therefore cannot predict the phase shift between
the shear velocity and the drag force.  We obtain this phase shift
from the experimental data of
\markcite{1989JFM...206..265J}{Jensen} {et~al.} (1989).

The thickness of the turbulent boundary layer ($d$) grows linearly
with the distance away from the edge of the plane ($\Delta \xi$). We
write
\begin{equation}
d = \lambda \Delta \xi,
\label{eq:dwithxi}
\end{equation}
where $\lambda$ is a small dimensionless number that describes the
expansion of the turbulent layer relative to the horizontal
displacement away from the edge.  If we define $v_*$ to be the root
mean squared fluctuating velocity in the turbulent layer, we find
$\lambda = v_*/\Delta v$

Within the boundary layer, the time-averaged velocity $u$ at distance
$y$ away from the plane obeys a logarithmic profile,
\begin{equation}
u(y)  = \frac{v_{*}}{c_1} \ln \left( \frac{c_2 v_{*} y}{\nu} \right),
\label{eq:logarithmicvelocity}
\end{equation}
where $c_1 \simeq 0.4$ and $c_2 \simeq 7.69$ are experimentally
determined constants. At $y=d$, we should recover the flow velocity,
so $\Delta v = v_* c_1^{-1} \ln(c_2 v_* d/\nu)$. Combining this
expression with equation \refnew{eq:dwithxi}, we find the following
solution for $\lambda$,
\begin{equation}
\frac{c_1}{\lambda}  =  \ln \left( c_2 \lambda^2\, \Rey \right),
\label{eq:lambda}
\end{equation}
where we define an experimentally controlled Reynolds number for the
flow by
\begin{equation}
\Rey \equiv  \frac{\Delta v \Delta \xi}{\nu}.
\end{equation}
A rough scaling for $\lambda$ can be found in the limit of very large
Reynolds number where the solution to equation \refnew{eq:lambda} is
\be
\lambda  \simeq  \frac{c_1}{\ln(c_2\,\Rey)}.
\label{eq:lambdasolution}
\ee
The fact that $\lambda$ is a slowly varying function of the Reynolds
number helps simplifying the calculation for energy dissipation rate.

The drag force per unit area on the surface due to the fluctuating
velocity field is
\begin{equation}
{\rm Drag} = \rho v_*^2 = \lambda^2 \rho \Delta v^2 \equiv 
\frac{1}{2} C_D \rho \Delta v^2 ,
\label{eq:dragold}
\end{equation}
which identifies the drag coefficient in the turbulent regime to be
$C_D= 2\lambda^2 \sim 2[c_1/\ln(c_2 \Rey)]^2$, also a slowly
decreasing function of the Reynolds number. 
\markcite{1989JFM...206..265J}{Jensen} {et~al.} (1989) 
show that in the fully turbulent regime, the drag force is nearly in
phase with the shear velocity.

The scalings $C_D=2 \Rey^{-1/2}$ in the laminar regime
\markcite{1959flme.book.....L}({Landau} \& {Lifshitz} 1959) and $C_D = 2 \lambda^2$ in the turbulent
regime compare very well with the experimental determinations of $C_D$
by \markcite{1989JFM...206..265J}{Jensen} {et~al.} (1989) over a large range of
Reynolds number.  Furthermore, if we estimate the critical Reynolds
number for the onset of turbulence by equaling the two $C_D$
expressions, we find $\Rey_{ \rm crit}=1.6 \times 10^5$, in good
agreement with the experiment. At $\Rey=\Rey_{crit}$, $\lambda=0.05$
and $C_D=0.005$. Beyond the onset of turbulence, $C_D$ decreases
logarithmically with the Reynolds number.

\section{Equilibrium Temperature Profile Around a Localized Source of Heat}
\label{S:ConductionCalc}

We assume thermal equilibrium has been reached. On either side of a
localized heat source, the heat conduction and energy conservation
equations read,
\begin{eqnarray}
\vec{F}  & = & - \kappa \vec{\nabla} T, \nonumber \\
\nabla\cdot {\vec{F}} & = & - \epsilon_\nu,
\label{eq:2diff}
\end{eqnarray}
where $\kappa$ is the thermal conductivity and scales inversely with
$T$ in the neutron star interior, while $\epsilon_\nu$ is the neutrino
emissivity with $\epsilon_\nu \propto T^8$ for modified Urca
reactions. We define two constants $\kappa_0 = \kappa T$ and
$\epsilon_0 = \epsilon_\nu/T^8$. 

Under the planar approximation, we manipulate the above equations to find
\begin{equation}
{{d\ln T}\over{dr}}\, {{d^2 \ln T}\over{dr^2}} = 
{{\epsilon_0}\over{\kappa_0}}\, T^8\, {{d\ln T}\over{dr}}.
\end{equation}
where $r$ is the radius. Integrating both sides, we find
\begin{equation}
\left. \left({{d\ln T}\over{dr}}\right)^2\right|^a_b = 
\left. {{\epsilon_0}\over{4 \kappa_0}} T^8\right|^a_b,
\end{equation}
where $a$ and $b$ are two arbitrary points within the integration
range.
Here, we take point $a$ to be at the boundary layer, which is the
source of heat, and point $b$ to be at many conduction lengths away
from the boundary layer so that the flux and temperature at point $b$
may be regarded as negligible. The flux and temperature at the
boundary are then related by
\begin{equation}
F = {1\over 2}(\epsilon_0 \kappa_0)^{1/2}\, T_{\rm bl}^4.
\end{equation}
If we define a conduction length $l = \kappa T_{\rm bl}/F =
\kappa_0/F$, we find $F = \epsilon_0\, l\, T_{\rm bl}^8/4$, a result
we adopt in \S \ref{sec:conduction}.

\section{Evolution Equations}
\label{S:evolutioneqn}

We have modified the phenomenological model of \markcite{Owenetal98}{Owen} {et~al.} (1998)
to calculate the effect of the r-mode on the spin and temperature of
the neutron star in the isothermal approximation.  The evolution
equations for mode amplitude $\alpha$, spin frequency $\Omega_s$, and
stellar temperature $T$ are
\begin{eqnarray}
\dot{\alpha} & = & - \frac{\alpha}{\tau_{\rm gr}} - \frac{\alpha}{\tau_v}
\frac{1-\alpha^2 Q}{1+\alpha^2 Q},
\label{eq:alphadot} \\
\dot{\Omega_s} & = & - \frac{2\Omega_s}{\tau_v} \frac{\alpha^2 Q}{1+\alpha^2 Q}
+ \frac{1}{\tilde{I}} \left( \frac{4}{3} \right)^{1/2} \frac{\dot{M}}{M} 
\Omega_d,
\label{eq:omegadot}\\
C_T T_9 \dot{T}_9 & = & - L_{\nu,9} T_9^8 
+ \epsilon_a (\dot{M}/M)(GM^2/2R)
+ 2E/\tau_{\rm heat},
\label{eq:Tdot}
\end{eqnarray}
where $Q=0.094$, $\tilde{I}=0.261$, $\dot{M}$ is the mass accretion
rate onto the surface of the neutron star (equals $0$ in the young
neutron star case), $C_T=1.4 \times 10^{48} \erg\s^{-1}$ is twice the
specific heat at a temperature of $10^9K$, and $L_{\nu,9}=7.4 \times
10^{39} \erg\s^{-1}$ is the neutrino cooling rate at $T=10^9 K$.  The
terms involving $\dot{M}$ describe the torque and heating by accretion
\markcite{1999ApJ...517..328L}({Levin} 1999), and $\epsilon_a$ is the fraction of 
accreted energy that goes into heating the star, we take it to be
unity. The mode energy $E = 0.5 \alpha^2 \Omega_s^2 MR^2 \tilde{J}$
with $\tilde{J}
\simeq 0.016$. The initial LMXB temperature is determined
self-consistently using equation \refnew{eq:Tdot} setting $\alpha = 0$
and $\dot{T} = 0$. It varies by a factor of a few over three orders of
magnitude in accretion rates.  The various time-scales in the above
equations are listed below.

Following \markcite{LindblomMendellOwen}{Lindblom} {et~al.} (1999), we write the energy input
rate due to gravitational wave back-reaction to be
\begin{equation}
\frac{1}{\tau_{\rm gr}} \equiv  - \frac{ \dot{E}_{\rm gr} }{2E} = 
\frac{1}{ \tilde{\tau}_{\rm gr}} \, \nukhz^6,
\label{eq:taugr}
\end{equation}
with $\tilde{\tau}_{\rm gr}=-18.7 \s$ and $\nukhz =
\nu_s/1\kHz$. The negative sign indicates driving.
Similarly, the bulk viscosity and molecular shear viscosity rates are
\begin{equation}
\frac{1}{\tau_{\rm bv}}  =  
\frac{1}{ \tilde{\tau}_{\rm bv} }\, \nukhz^2\, T_8^6, {~~}{\rm and}{~~}
\frac{1}{\tau_{\rm sv}} =
\frac{1}{ \tilde{\tau}_{\rm sv} }\, T_8^{-2},
\label{eq:2rates}
\end{equation}
with $\tilde{\tau}_{\rm bv}= 3.5\times 10^{17} \s$ and
$\tilde{\tau}_{sv}=2.5 \times 10^6 \s$. We will continue to use these
estimates in the presence of a crust even though the r-mode occupies a
smaller volume.

BU and this article have pointed to the importance of the boundary
layers 
at the core-crust interface. We model the dissipation rate in
this layer as a function that is continuous when the layer changes
from laminar (viscous) to turbulent, \be \frac{1}{\tau_{\rm bl}} =
\Theta(\alpha_{\rm eq}-\alpha)\, \frac{1}{\tau_{\rm vbl}} +
\Theta(\alpha-\alpha_{\rm eq})\, \frac{1}{\tau_{\rm tbl}},
\label{eq:tau_bl}
\ee
with $\Theta(x) = 1$ when $x > 0$ and $0$ otherwise. The viscous and
the turbulent damping rates are, respectively,
\begin{equation}
\frac{1}{\tau_{\rm vbl}}  =  \frac{1}{ \tilde{\tau}_{\rm vbl} }\, \eta^2
\,\nukhz^{1/2}\, T_8^{-1} {~~~~~~}{\rm and}{~~~~~~}
\frac{1}{\tau_{\rm tbl}} = \frac{1}{ \tilde{\tau}_{\rm tbl} }\,
\eta^3\, C_D\, \alpha\, \nukhz,
\end{equation}
where the fiducial time-scales $\tilde{\tau}_{\rm vbl}=65 \s$ and
$\tilde{\tau}_{\rm tbl}=3.3 \times 10^{-4}s$.  We equate $\tau_{\rm
vbl}$ with $\tau_{\rm tbl}$ to obtain the swith-over amplitude
$\alpha_{\rm eq} = 1.0 \times 10^{-3}\, \eta^{-1}\, \nukhz^{-1/2}\,
T_8^{-1}$.  In our calculations, we evaluate $C_D$ using the Reynolds
number at the equator.

The total rate of viscous damping for the r-mode as well as the rate
of heating due to the r-mode (ignoring bulk viscosity heating) are
given by
\begin{equation}
\frac{1}{\tau_v} = \frac{1}{\tau_{\rm bv}} + \frac{1}{\tau_{\rm sv}}
+ \frac{1}{\tau_{\rm bl}} {~~~~~~}{\rm and}{~~~~~~}
\frac{1}{\tau_{\rm heat}} \simeq \frac{1}{\tau_{\rm sv}}
+ \frac{1}{\tau_{\rm bl}}.
\label{eq:tauheat}
\end{equation}

\end{document}